\documentclass[aps,prfluids]{revtex4-2}
\usepackage{hyperref}
\usepackage{graphicx} 
\usepackage{xcolor} 

\newcommand{\Ra}{\mathrm{Ra}}
\newcommand{\Ek}{\mathrm{Ek}}
\newcommand{\Nu}{\mathrm{Nu}}
\newcommand{\Fr}{\mathrm{Fr}}
\newcommand{\Ro}{\mathrm{Ro}}
\renewcommand{\Pr}{\mathrm{Pr}}

\newcommand{\red}[1]{\textcolor{black}{#1}}

\begin{document}
\title{Experiments on \red{rapidly rotating} convection: the role of the Prandtl number}
\date{\today}

\author{Hannah M. Clercx}
\altaffiliation{Present address: Environmental Fluid Mechanics group at Faculty of Civil Engineering, Delft University of Technology, P.O. Box 5, 2600 AA Delft, The Netherlands}\affiliation{Fluids and Flows group and J.M. Burgers Center for Fluid Mechanics, Department of Applied Physics and Science Education, Eindhoven University of Technology, P.O. Box 513, 5600 MB Eindhoven, The Netherlands}
\author{Rudie P.J. Kunnen}\email[Contact author: ]{r.p.j.kunnen@tue.nl}
\affiliation{Fluids and Flows group and J.M. Burgers Center for Fluid Mechanics, Department of Applied Physics and Science Education, Eindhoven University of Technology, P.O. Box 513, 5600 MB Eindhoven, The Netherlands}

\begin{abstract}
    \noindent Flows at planetary scales are generally driven by buoyancy and influenced by rotation. Rotating Rayleigh-Bénard convection (RRBC) is a practical and simple model that can be used to describe these systems. In RRBC, thermally induced convection occurs, which is influenced by the constant rotation it experiences. We study RRBC in a cylinder in the \red{transition region between rotation-affected and rotation-dominated (also called geostrophic) convection}. 
    Experiments are performed to assess the dependence of the Nusselt number $\Nu$ (efficiency of convective heat transfer) on the Prandtl number $\Pr$ (ratio of kinematic viscosity over thermal diffusivity), a relation that is not explored much for geostrophic convection. By using water at different mean temperatures we can reach $2.8\le \Pr\le 6$. We study the relation between $\Pr$ and $\Nu$ at constant Ekman number $\Ek=3\times10^{-7}$ (an inverse measure for strength of rotation) for two different diameter-to-height aspect ratios ($\Gamma=1/5$ and $1/2$) of the setup. The corresponding constant Rayleigh numbers (strength of thermal forcing) are $\Ra=1.1\times 10^{12}$ and $1\times 10^{11}$, respectively. Additionally, we measure the relation between the Rayleigh number $\Ra$ and $\Nu$ for $4\times10^{10}\le \Ra\le 7\times10^{11}$, $\Ek=3\times10^{-7}$ and $\Pr=3.7$. It is found that $\Nu$ exhibits a significant dependence on $\Pr$, even within this limited range. Increasing $\Pr$ by a factor 2 resulted in a decrease of $\Nu$ of about $25 \%$. We hypothesize that the decrease of $\Nu$ is caused by the changing ratio of the thermal and kinetic boundary layer thicknesses as a result of increasing $\Pr$. We also consider the anticipated contributions of the wall mode to the heat transfer using sidewall temperature measurements.
\end{abstract}

\maketitle

\section{Introduction\label{ch:intro}}
Rotating Rayleigh-Bénard convection (RRBC) is a simplified model intended to represent geophysical and astrophysical flows, as it considers flows driven by buoyancy and affected by rotation. In this model, a flow is confined between two horizontal plates, heated from below and cooled at the top, while rotating about a vertical axis. Regimes in RRBC vary from the rotation-unaffected regime to the rotation-dominated (or geostrophic) regime \cite{Kunnen2021,Ecke2023}. These regimes can be defined using dimensionless parameters; here we use the Rayleigh number $\Ra$, the Ekman number $\Ek$ and the Prandtl number $\Pr$. We define 
    \begin{equation}
        \Ra=\frac{g\alpha}{\nu\kappa}\Delta T H^3 \, , \quad \Ek=\frac{\nu}{2\Omega H^2} \, , \quad \Pr=\frac{\nu}{\kappa} \, ,
    \end{equation}
where $g$ is gravitational acceleration, $H$ the distance between the plates and $\Delta T$ the temperature difference between them, $\Omega$ the rotation rate and $\alpha$, $\nu$ and $\kappa$ are the thermal expansion coefficient, kinematic viscosity and thermal diffusivity of the fluid, respectively. $\Ra$ is an indication for the strength of thermal forcing, $\Ek$ quantifies the ratio between the viscous and the Coriolis forces, $\Pr$ quantifies the ratio between the momentum diffusivity and the thermal diffusivity. As a result, it is related to the ratio of the kinetic boundary layer and the thermal boundary layer. When $\Pr>1$, the kinetic boundary layer is expected to be thicker than the thermal boundary layer, whereas when $\Pr<1$, the reverse is true \cite{Schlichting1979}. Another parameter that is frequently encountered in the literature is the convective Rossby number $\Ro$, defined as \cite{Kunnen2021,Ecke2023}
\begin{equation}\label{eq:Ro}    \Ro=\frac{\sqrt{g\alpha\Delta T H}}{2\Omega H}=\Ek\sqrt{\frac{\Ra}{\Pr}} \, ,
\end{equation}
giving an \emph{a priori} indication of the ratio of the strength of thermal forcing to the importance of rotation.

\red{When rotation is applied to thermal convection, a larger temperature difference across the fluid layer (i.e. a larger Rayleigh number) is required before convection sets in. Using linear stability analysis, Chandrasekhar \cite{Chandrasekhar1961} found the following asymptotic relation for the critical Rayleigh number $\Ra_c$ (valid for small $\Ek$):
\begin{equation}
    Ra_c=8.70\Ek^{-4/3} \, .
\end{equation}
This relation provides us with another useful parameter $\Ra/\Ra_c$, expressing the degree of supercriticality. A related parameter $\widetilde{\Ra}=\Ra\Ek^{4/3}=8.70\Ra/\Ra_c$ is often employed in asymptotic studies (e.g., \cite{Julien2012,Maffei2021}), where the limit $\Ek\ll 1$ is applied and $\widetilde{\Ra}$ is the finite-valued aggregate input parameter.}

Recently, the geostrophic regime of \red{rapidly} rotating convection (or geostrophic convection in short) has been studied more extensively, as it is believed that this regime most accurately represents geophysical and astrophysical flows \cite{Kunnen2021}. The geostrophic regime is characterized by small Ekman and Rossby numbers, resulting in a principal balance of forces between the Coriolis force and the pressure gradient: the so-called geostrophic balance. As a result the flow is rotation-dominated: the flow is organized along the rotation axis but with turbulent fluctuations remaining (often referred to as quasi-two-dimensional turbulence, e.g. \cite{Alexakis2018,Kunnen2021}).

It is a challenge to reach the geostrophic regime in current experimental setups. Large-scale setups are required to simultaneously achieve large enough Rayleigh numbers to induce bulk convection and small enough Ekman numbers for rotational constraint \cite{Cheng2018}. This leaves many open questions concerning the dependency of the heat transfer on certain parameters \red{within the transition range between rotation-affected and geostrophic convection \cite{Kunnen2021}}. With this in mind, an experimental study is done on the variation of the Prandtl number $\Pr$. We want to determine whether and how this affects the efficiency of convective heat transfer, parameterized by the Nusselt number
\begin{equation}\label{eq:Nu}
    \Nu=\frac{qH}{k\Delta T} \, ,
\end{equation}
where $q$ is the heat flux and $k$ the thermal conductivity.

The heat flux in 
\red{rapidly} rotating convection has only scarcely been probed with experiments. Most studies have used constant $\Pr$ with water ($\Pr\approx4-7$) \cite{Cheng2015, Cheng2020, Lu2021, Hawkins2023} or gases ($\Pr\approx0.7$) \cite{Ecke2014, Wedi2021} as the working fluid. Experimental work on rotating convection with variation of $\Pr$ has been done, but mostly outside of the geostrophic regime at higher values of $\Ek\gtrsim 10^{-6}$ \cite{Liu1997,Zhong2009,Weiss2016}. Abbate \& Aurnou \cite{Abbate2023} could reach a smaller $\Ek\ge 2\times 10^{-7}$ at $\Ra$ up to $2\times 10^{12}$, employing water and other liquids to have $\Pr$ between $6$ and $10^3$. Direct Numerical Simulation (DNS) studies investigating variation of $\Pr$ in rotating convection have used cylindrical domains \cite{Zhong2009,Stevens2010,Zhang2021} or rectilinear domains with periodic boundary conditions in the horizontal directions \cite{King2013,Stellmach2014,Yang2020,AguirreGuzman2022,Anas2024}. Additionally, effects of $\Pr$ have been considered in simulations employing an asymptotically reduced set of governing equations valid in the limit of rapid rotation \cite{Sprague2006,Julien2012,Maffei2021}. These simulations also consider a horizontally periodic domain. We summarize the parameter ranges for these studies in Table \ref{tab:literature}. The main conclusion from these works is that, for larger $\Pr\gtrsim 1$ and at moderate values of $\Ra\lesssim 10^{10}$ and $\Ek\gtrsim 10^{-6}$, there exists a range of parameter values where the convective heat transfer is larger with rotation than without. \red{This range is bounded from above by a geometry-dependent convective Rossby number \cite{Weiss2010}}
\begin{equation}
    \red{\Ro=\frac{a}{\Gamma}\left(1+\frac{b}{\Gamma}\right)\quad\mathrm{with}\quad a=0.381, b=0.061,}
    \label{eq:weiss2010}
\end{equation}
\red{where $\Gamma=D/H$ is the diameter-to-height aspect ratio for confined cylindrical domains.} The excess \red{heat transfer compared to the nonrotating case} (can be up to about $60\%$ \cite{Anas2024}) generally increases with $\Pr$ and reaches its maximum at progressively higher rotation rates (smaller $\Ek$) \cite{Stevens2010,Yang2020,Anas2024}. The mechanism involved is Ekman pumping \cite{Pedlosky1982,Kundu2008} acting as an efficient transport mechanism for boundary-layer fluid towards the opposite plate, where an `optimal' condition is found when kinetic and temperature boundary layers are of equal thickness. However, when moving to more extreme parameters $\Ra\gtrsim 10^{10}$ and $\Ek\lesssim 10^{-6}$, the convective heat flux overshoot with respect to the non-rotating case vanishes \cite{Cheng2015,Cheng2020}.

\begin{table}
    \caption{\label{tab:literature}Parameter overview of studies of turbulent rotating convection with variation of the Prandtl number $\Pr$. Note that the asymptotic simulations do not separately consider $\Ra$ and $\Ek$ as input parameters. Instead, they employ $\widetilde{\Ra}=\Ra\Ek^{4/3}$.}
    \begin{tabular}{llcccl}
        \hline
        \hline
        Method & Authors & $\Ra$ range & $\Ek$ range & $\Pr$ range and working fluid \\
        \hline
        Experiment & Liu \& Ecke \cite{Liu1997} & $2\times 10^5-5\times 10^8$ & $10^{-5}-\infty$ & $3-7$ (water) \\
                   & Zhong et al. \cite{Zhong2009} & $3\times 10^8-2\times 10^{10}$ & $5\times 10^{-6}-\infty$ & $3-6.5$ (water) \\
                   & Weiss et al. \cite{Weiss2016} & $4\times 10^8-4\times 10^{11}$ & $4\times 10^{-6}-4\times 10^{-3}$ & $0.74$ (N$_2$), $0.84$ (SF$_6$), $3-6.4$ (water), \\
                   & & & & $12.34$ (FC72), $24-36$ (isopropanol) \\
                   & Abbate \& Aurnou \cite{Abbate2023} & $3\times 10^8-2\times 10^{12}$ & $2\times 10^{-7}-4\times 10^{-4}$ & $6$ (water), \\
                   & & & & $41,206,993$ (3, 20, 100 cSt silicone oil) \\
                   & This work & $4\times 10^{10}-7\times 10^{11}$ & $3\times 10^{-7}$ & $2.8-6$ (water) \\
        \hline
        Cylinder DNS & Zhong et al. \cite{Zhong2009} & $10^8$ & $5\times 10^{-6}-\infty$ & $0.7-20$ \\
                     & Stevens et al. \cite{Stevens2010} & $10^8$ & $9\times 10^{-6}-\infty$ & $0.7-55$ \\
                     & Zhang et al. \cite{Zhang2021} & $5\times 10^7-5\times 10^9$ & $9\times 10^{-7}-4\times 10^{-5}$ & $0.1-12.3$ \\
        \hline
        Periodic DNS & King et al. \cite{King2013} & $10^3-10^9$ & $10^{-6}-\infty$ & $1,7,100$ \\
                     & Stellmach et al. \cite{Stellmach2014} & $2\times 10^{10}-2\times 10^{11}$ & $10^{-7}$ & $1,3,7$ \\
                     & Yang et al. \cite{Yang2020} & $10^7-2\times 10^9$ & $10^{-6}-\infty$ & $4.38, 6.4, 25, 100$ \\
                     & Aguirre Guzm\'an et al. \cite{AguirreGuzman2022} & $10^{10}-3\times 10^{12}$ & $10^{-7}-6\times 10^{-6}$ & $0.1, 5.2, 5.5, 100$ \\
                     & Anas \& Joshi \cite{Anas2024} & $2\times 10^4-2\times 10^{10}$ & $7\times 10^{-7}-\infty$ & $1-1000$ \\
        \hline
        Asymptotics & Sprague et al. \cite{Sprague2006} & \multicolumn{2}{c}{$20\le\widetilde{\Ra}\le 160$} & $1,7,\infty$ \\
                    & Julien et al. \cite{Julien2012} & \multicolumn{2}{c}{$10\le\widetilde{\Ra}\le 160$} & $1,3,7,15,\infty$ \\
                    & Maffei et al. \cite{Maffei2021} & \multicolumn{2}{c}{$20\le\widetilde{\Ra}\le 200$} & $1, 1.5, 2, 2.5, 3, 7$ \\
        \hline
        \hline
    \end{tabular}
\end{table}

Abbate \& Aurnou \cite{Abbate2023} have covered \red{rapidly rotating} convection with large $\Pr\ge 6$. But a large gap in $\Pr$ exists between these results and the results using gases ($\Pr\approx 0.7$). In this paper, we attempt to bridge the gap between these media. Inspired by earlier works who changed $\Pr$ by using water at different operating temperatures \cite{Liu1997,Zhong2009}, we use water at mean temperatures between $26$ and $61^\circ$C to achieve $2.8\le \Pr\le 6$. We measure the dependence of the Nusselt number $\Nu$ on the Prandtl number $\Pr$ at a constant Ekman number $\Ek=3\times 10^{-7}$. Additionally, we consider a $\Nu(\Ra)$ scan at constant $\Pr=3.7$, which is considerably lower than our earlier experiments at $\Pr\approx 5$ \cite{Cheng2020}.

\begin{figure}
    \centerline{\raisebox{0.37\textwidth}{(a)}\includegraphics[scale=0.57]{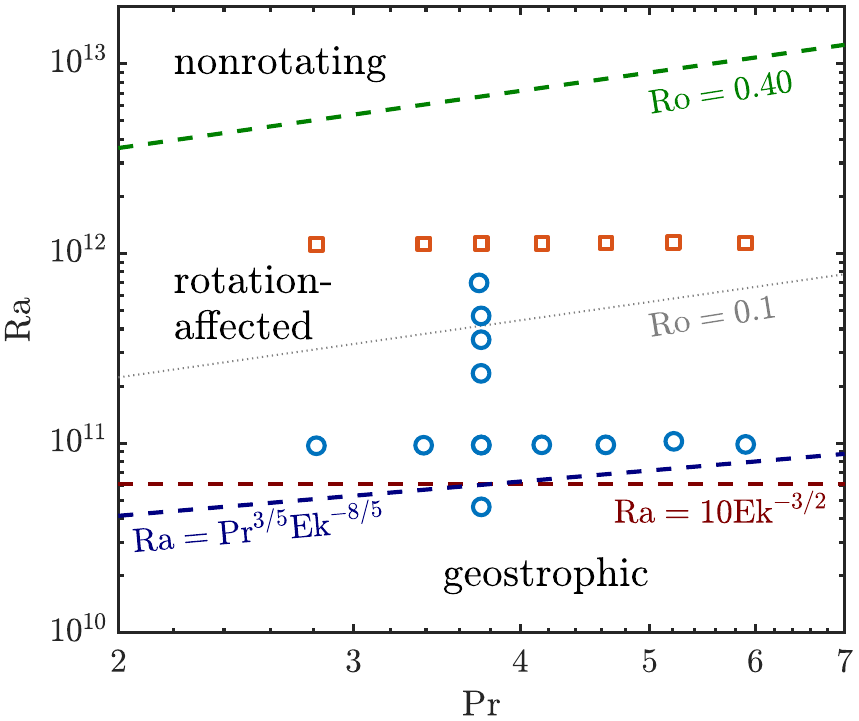}}
    \centerline{\raisebox{0.37\textwidth}{(b)}\includegraphics[scale=0.57]{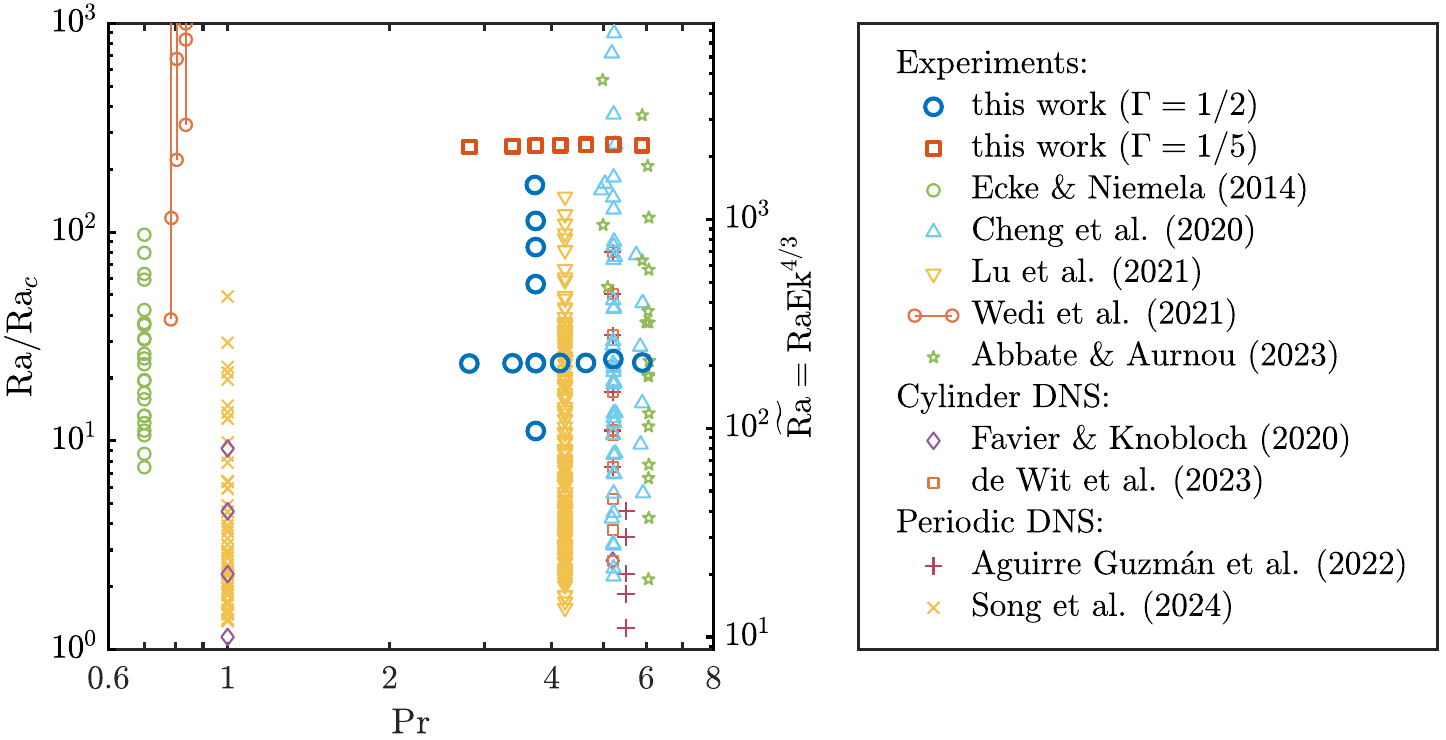}}
    \caption{Overview of conducted experiments. \red{All our data points are at $\Ek=3\times 10^{-7}$. (a) $(\Pr,\Ra)$ diagram with regime boundaries from literature.} Blue circles indicate the setup with $\Gamma=1/2$ while red squares indicate the setup with $\Gamma=1/5$. The measurements for $\Gamma=1/2$ were on a range of $\Pr$ at constant $\Ra=1\times 10^{11}$ as well as on a range of $\Ra$ at constant $\Pr=3.7$. The measurements for $\Gamma=1/5$ were on the same range of $\Pr$ but at $\Ra=1.1\times 10^{12}$. \red{The transition between the nonrotating and rotation-affected range (green dashed line) is based on the aspect-ratio-dependent criterion (\ref{eq:weiss2010}) due to Weiss et al. \cite{Weiss2010}, which at $\Gamma=1/5$ gives $\Ro=0.40$. Two suggested relations for the transition between the rotation-affected and geostrophic range are included: $\Ra=10\Ek^{-3/2}$ due to King et al. \cite{King2012,King2013} (red dashed line) and $\Ra=\Pr^{3/5}\Ek^{-8/5}$ from Cheng et al. \cite{Cheng2020} (blue dashed line).} The \red{dotted line} indicate\red{s $\Ro=0.1$} for reference. \red{(b) $(\Pr,\Ra/\Ra_c)$ diagram with a comparison of other literature data for rapidly rotating convection ($\Ek\le 10^{-6}$). Apart form our current data, we include the experimental works of Ecke \& Niemela (2014) \cite{Ecke2014}, Cheng et al. (2020) \cite{Cheng2020}, Lu et al. (2021) \cite{Lu2021}, Wedi et al. (2021) \cite{Wedi2021}, Abbate \& Aurnou (2023) \cite{Abbate2023}; the cylinder DNS works of Favier \& Knobloch (2020) \cite{Favier2020}, de Wit et al. (2023) \cite{DeWit2023}; and the periodic DNS works of Aguirre Guzm\'an et al. (2022) \cite{AguirreGuzman2022}, Song et al. (2024) \cite{Song2024scaling,Song2024}. The corresponding $\widetilde{\Ra}=\Ra\Ek^{4/3}$ values are indicated (right ordinate).}}
    \label{fig:PhaseDiagram}
\end{figure}

\red{To put our new data in context, we plot our data points in phase diagrams in Fig. \ref{fig:PhaseDiagram} along with regime transition relations and reference data from the literature. Panel (a) displays the current experiment settings in the $(\Pr,\Ra)$ parameter space, with regime transitions between nonrotating, rotation-affected and rotation-dominated (geostrophic) convection \cite{Weiss2010,King2012,Cheng2020,Kunnen2021}. Panel (b) plots our points on the $(\Pr,\Ra/\Ra_c)$ parameter space along with reference data for rapidly rotating convection, i.e. we restrict ourselves to data for which $\Ek\le 10^{-6}$. In that graph, we also include values of $\widetilde{\Ra}$ to allow an indicative comparison with the asymptotic studies \cite{Sprague2006,Julien2012,Maffei2021}.}

One important aspect of confined rotating convection is the occurrence of a so-called wall mode: a strong, traveling wave of up- and downward flow forming near the sidewall that persists into the turbulent regime and contributes strongly to the overall convective heat transfer \cite{DeWit2020,Zhang2020,Favier2020,Zhang2021,DeWit2023}. This complicates the interpretation of global measurements like the Nusselt number and the comparison with numerical simulation results from horizontally periodic domains without a sidewall. We evaluate the amplitude of the wall mode from temperature measurements at the sidewall and discuss expected influences of the wall modes on the current measurements.

\section{Methodology\label{ch:meth}}
TROCONVEX is an experimental setup designed for rotating Rayleigh-Bénard convection experiments in the geostrophic regime \cite{Kunnen2021}. The apparatus is located at Eindhoven University of Technology. The setup has been extensively described in Refs. \cite{Cheng2018,Cheng2020,Madonia2022}; here we summarize the most important points and indicate the relevant differences in operation. The setup consists of stackable cylinders made of Lexan, which can be built up to a total height $H=4$ m. However, in the current experiments, we only used $H=0.8$ and $2$ m. The diameter $D$ is 0.39 m. The top and bottom are closed with copper plates that are cooled and heated, respectively. The working fluid in the cylinder is water. The mean temperature $T_m$ and the temperature difference $\Delta T$ are calculated using the average of the temperatures $T_b$ of the bottom plate and $T_t$ of the top plate: $T_m=(T_t+T_b)/2$ and $\Delta T=T_b-T_t$. The bottom plate is heated through the use of an electrical resistance heater and its temperature $T_b$ is controlled towards a desired setpoint. Similarly, the top plate is cooled by circulating coolant from a refrigerated bath, which is also controlled towards a setpoint $T_t$. The setup is thermally insulated with foam and heat shields. Below the bottom plate, there is a second heated plate that is controlled to the same temperature $T_b$, thereby reducing conductive heat losses. Similarly, an arrangement of sidewall heat shields (5 in total for $H=2$ m) surrounds the cylinder laterally. Each segment takes on the mean temperature measured at the corresponding height in the sidewall, thereby reducing lateral conductive heat losses. The procedure to quantify statistical accuracy of the measurements is treated in detail in Ref. \cite{Cheng2020}. In the current paper, we write error intervals for any quantity $x$ as $x\pm\sigma_x$, i.e. $\pm$ one standard deviation $\sigma_x$.

In the current experiments, the mean temperature $T_m$ of the water is varied to obtain a range of Prandtl numbers $\Pr$. This is achieved by setting appropriate values for the temperatures of the top and bottom plates. The range of $\Pr$ that can be reached depends on the specifications of the setup. The material that is used for the cylinders, Lexan, can withstand temperatures up to 120$^\circ$C without degradation. The maximal mean temperature that we use is $T_m=61^\circ$C. Note that the bottom temperature will then be higher; in practice the peripheral equipment (heating and cooling, as well as materials in the setup) could manage a highest $T_b\approx 63^\circ$C. The lowest $T_m$ was limited by the cooling capacity of the cooling system against the room temperature and temperature differences, resulting in a lowest $T_m$ of $26^\circ$C. For the measurements at varying $\Ra$, $T_m$ is kept constant while the temperature difference $\Delta T$ is varied, again by setting appropriate values for the plate temperatures. \red{In each experiment reported in this paper we set the rotation rate $\Omega$ such that a constant Ekman number $\Ek=3\times 10^{-7}$ is maintained.}

A \red{graphical} overview of the experiments that have been performed can be seen in Figure \ref{fig:PhaseDiagram}; tables with the numerical parameter values for all experiments can be found in Appendix \ref{app:experiments}. There, for completeness, we also plot the dependence on temperature of the fluid properties in Figure \ref{fig:waterproperties}. The full dataset is also available \cite{dataset}.

Some practical points concerning operation of the experiment must be mentioned. To be able to change $\Ra$ sufficiently, we used different heights $H=0.8$ and $2$ m (with corresponding aspect ratios $\Gamma=0.494$ and $0.195$, respectively, referred to in this paper as $1/2$ and $1/5$). The characteristic thickness of the wall mode scales as $\delta_S\sim \Ek^{1/3}H$ \cite{DeWit2020,Zhang2021}. Assuming a prefactor of $1$, a reasonable choice based on earlier work \cite{Kunnen2010,Kunnen2011,DeWit2020}, $\delta_S=5.4$ mm for $\Gamma=1/2$ while $\delta_S=13.4$ mm for $\Gamma=1/5$. So the wall mode covers a larger fraction of the total volume of the larger cylinder. This could lead to a $\Gamma$-dependent heat transfer. Additionally, the convective heat flux carried by the wall mode is expected to be larger at smaller $\Pr$ \cite{Zhang2021}. This property is considered in the current experiments with measurements of temperature in the sidewall. Furthermore, to ensure that the influence of centrifugal buoyancy remains small, we keep the Froude number $\Fr=\Omega^2D/(2g)$ (ratio of centrifugal acceleration at the sidewall to gravitational acceleration) at a sufficiently low value. This was taken into account during all experiments. The highest $\Fr$ value reached was $\Fr=0.099$, a value that we found in our previous work \cite{Cheng2020} to be low enough to satisfactorily retain up-down symmetry in the mean temperature profiles.

\section{Results and discussion\label{ch:res}}
\subsection{Heat flux at constant $\Ra$}
\begin{figure}
    \begin{tabular}[t]{llll}
    \raisebox{0.37\textwidth}{(a)} & \includegraphics[scale=0.57]{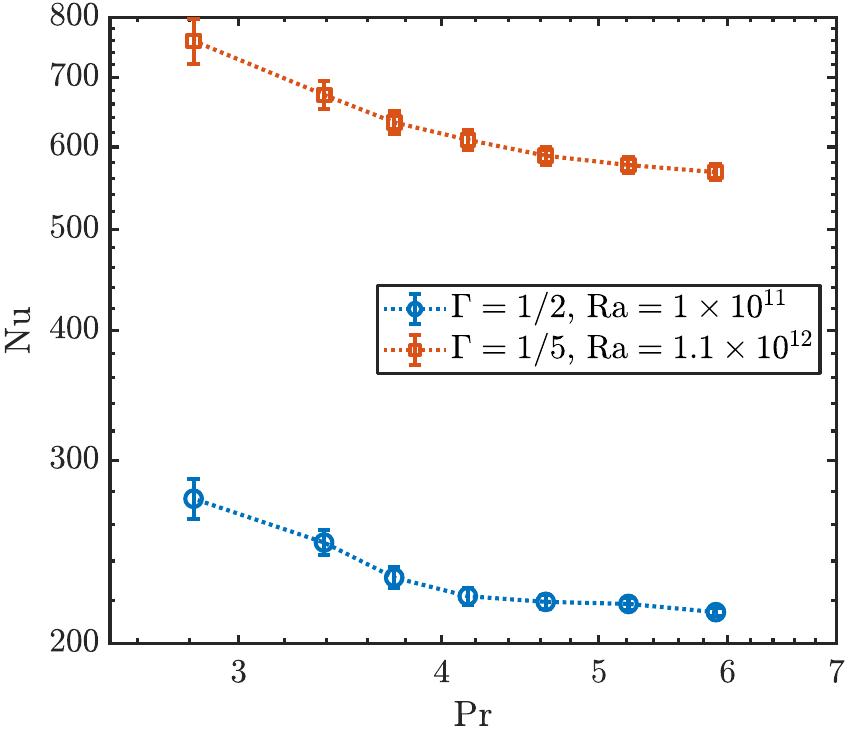} &
    \raisebox{0.37\textwidth}{(b)} & \includegraphics[scale=0.57]{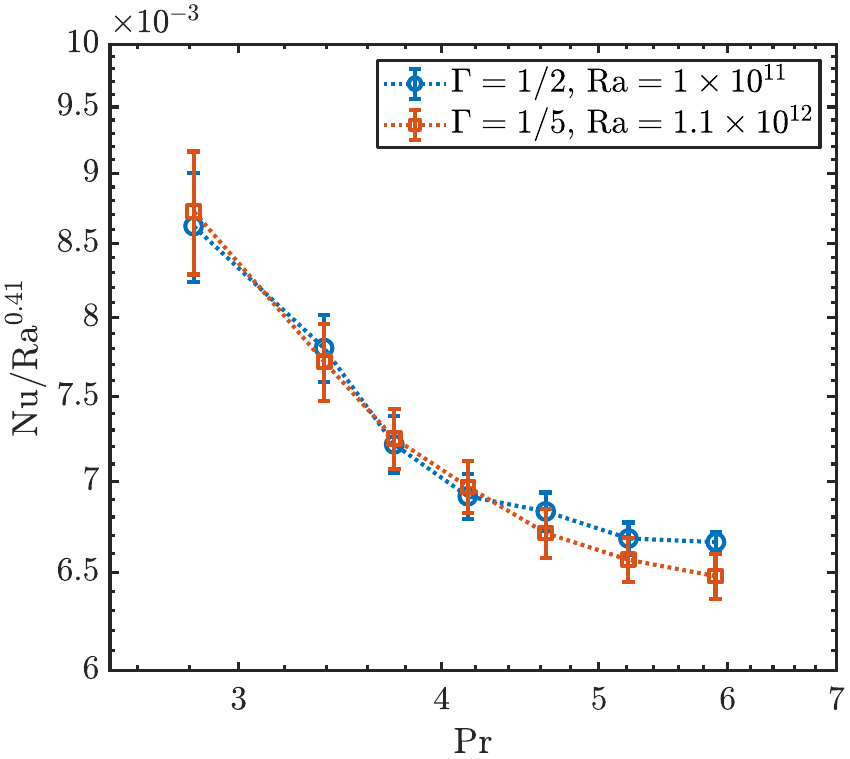}
    \end{tabular}
    \caption{(a) Plot of $\Nu$ as a function of $\Pr$ for $\Gamma=1/2$ (blue, $\Ra = 1\times10^{11}$, \red{Ek$=3.1\times10^{-7}$, Ro=$0.040-0.056$}) and $\Gamma=1/5$ (red, $\Ra = 1.1\times10^{12}$, \red{Ek$=3\times10^{-7}$, Ro=$0.131-0.189$}). (b) The same data compensated by the empirical factor $\Ra^{0.41}$. \red{Note that in order to keep Ek constant, $\Omega$ was decreased as $\Pr$ decreased.}}
    \label{fig:Nu_vs_Pr}
\end{figure}

We start with the two measurement series of $\Nu$ as a function of $\Pr$ at constant $\Ra$. Figure \ref{fig:Nu_vs_Pr}(a) shows the experimental data for both series at different aspect ratio $\Gamma$ \red{(with $\Ek=3.10\times10^{-7}$ for $\Gamma=1/2$ and $\Ek=3.00\times10^{-7}$ for $\Gamma=1/5$)}. These results clearly show that changing $\Pr$ has a significant effect on $\Nu$. Increasing $\Pr$ from $2.8$ to $6$ results in a gradually downward trend that is steepest at the lowest $\Pr=2.8$. Due to the difference in $\Ra$ of about a factor $10$, the values of $\Nu$ differ significantly between the two sets. However, when compensating with the empirically obtained scaling factor $\Ra^{0.41}$, see figure \ref{fig:Nu_vs_Pr}(b), we can see a rather satisfactory collapse. From this we conclude that\red{, in contrast to non-rotating convection \cite{Ahlers2022}, under rapid rotation the} aspect ratio $\Gamma$ does not appear to influence the convective heat flux directly. However, the wall mode could cause some differences given the different volume fractions it covers, a topic we will discuss in more detail in Sec. \ref{ch:wallmode}. We also note that the scaling factor $\Ra^{0.41}$ represents a steeper $\Ra$ dependence here than for non-rotating experiments, where exponents just under $1/3$ are usually reported ($\Ra^{0.308}$ from our previous measurements \cite{Cheng2020}).
 
For the interpretation of these data, we invoke the arguments put forth by Stevens et al. \cite{Stevens2010}. The ratio of the thicknesses of kinetic and thermal boundary layers (BLs) is expected to play a decisive role for the efficiency of convective heat transfer. For $\Pr\ll 1$, the kinetic BL is expected to be thinner than the thermal BL, whereas for $\Pr\gg 1$ the opposite is true. At some intermediate $\Pr$, these boundary layers will be (roughly) equal. This couples to the efficiency of convective heat transfer $\Nu$ involving the efficiency of Ekman pumping \cite{Pedlosky1982,Kundu2008}. If the kinetic BL is much thicker than the thermal BL ($\Pr\gg 1$), the pumping is inefficient as only a small fraction of the Ekman-pumped fluid is from within the thermal BL (with strong temperature contrast relative to the bulk), resulting in a comparatively lower value of $\Nu$. Likewise, if the thermal BL is thicker than the kinetic BL ($\Pr\ll 1$), the Ekman-pumping-fed vortical plumes will rapidly lose coherence due to efficient diffusion of heat from the plumes, again resulting in a lower value of $\Nu$. In the region where the boundary layers are of similar thickness, Ekman pumping is most efficient at transporting fluid from within the thermal BL towards the vertically opposite side, resulting in an optimum $\Nu$. Based on these measurements, we expect the optimum to be found at $\Pr$ lower than our minimal value of $2.8$ where we cover the downslope from the peak. Unfortunately, we cannot directly measure BL thicknesses in this setup.

An alternative argumentation uses the convective Rossby number $\Ro$. By considering measurements at different $\Pr$ while keeping $\Ra$ and $\Ek$ constant, we do change the value of $\Ro$, see Eq. (\ref{eq:Ro}). For $2.8\le\Pr\le6$, we cover a range of $0.056\ge\Ro\ge 0.040$ for $\Ra=1\times 10^{11}$ and $0.19\ge\Ro\ge 0.13$ for $\Ra=1.1\times 10^{12}$. So the high-$\Pr$ cases are at lower $\Ro$, meaning that a slightly stronger rotational constraint is expected there which suppresses convective heat transfer. For reference, Weiss \& Ahlers \cite{Weiss2011} measured $\Nu$ as a function of $\Ek$ (or $\Ro$) at constant $\Ra=7.2\times 10^{10}$ and $\Pr=4.38$. They report maximal $\Nu$ at $\Ro=0.35$ with a decrease for lower $\Ro$, a trend consistent with our current results for $\Ro\le0.19$.
 
\subsection{Heat flux at constant $\Pr$}
\begin{figure}
    \begin{tabular}[t]{llll}
    \raisebox{0.37\textwidth}{(a)} & \includegraphics[scale=0.57]{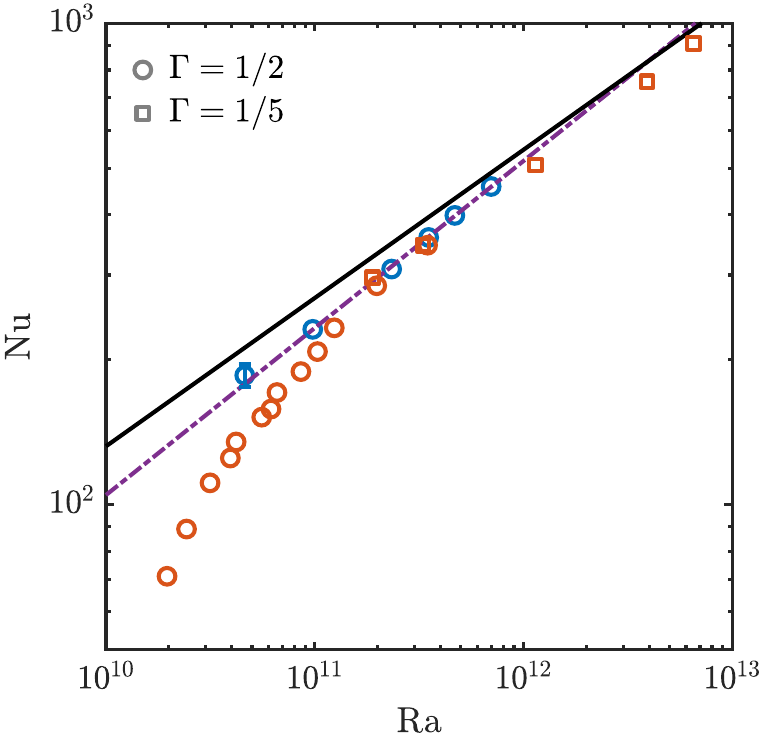} &
    \raisebox{0.37\textwidth}{(b)} & \includegraphics[scale=0.57]{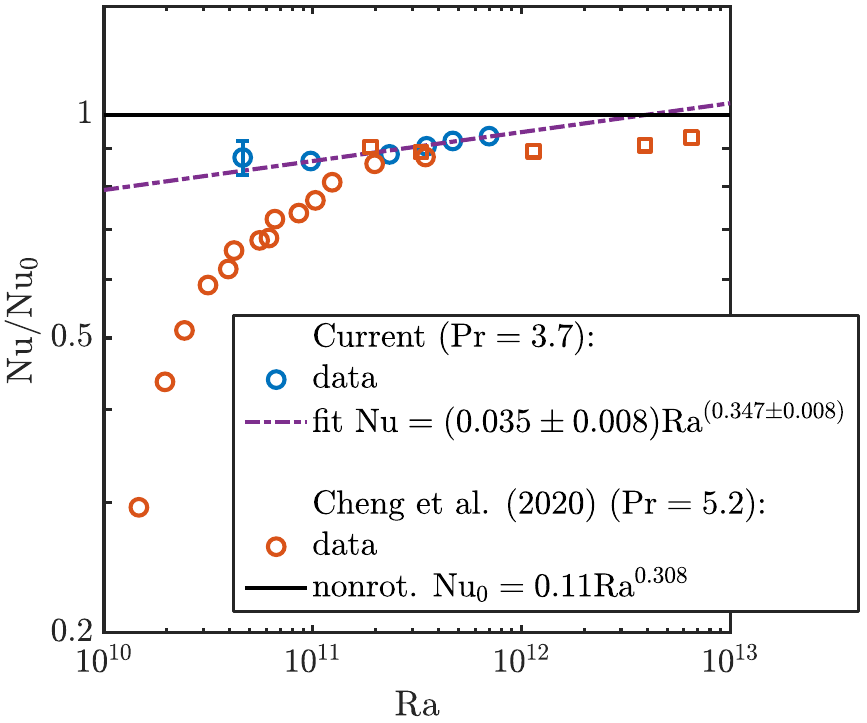}
    \end{tabular}
    \caption{\red{(a)} Scaling of $\Nu$ as a function of $\Ra$. The blue markers indicate the current experimental data, with \red{Pr$=3.7$ and Ek$=3\times10^{-7}$ ($\Ro=0.035-0.134$) and} the purple dashed line indicating the fit of the relation $\Nu(\Ra)$. The red markers indicate data from Cheng et al. \cite{Cheng2020}, which is from experiments on the same setup but at higher \red{$\Pr=5.2$ and at the same $\Ek=3\times10^{-7}$}. \red{Circles indicate $\Gamma=1/2$ and squares indicate $\Gamma=1/5$.} The black line indicates the relation between $\Nu_{\red{0}}$ and $\Ra$ in non-rotating convection. \red{(b) The same data, normalized with the nonrotating $\Nu_0$ result from Cheng et al. \cite{Cheng2020}}.}
    \label{fig:Nu_vs_Ra}
\end{figure}

We next measure $\Nu(\Ra)$ at lower $\Pr=3.7$ than in our previous study in the same setup \cite{Cheng2020}. The results are plotted in Figure \ref{fig:Nu_vs_Ra}\red{(a)}, with the $\Pr=5.2$ data at the same $\Ek=3\times 10^{-7}$ from Ref. \cite{Cheng2020} included for reference. The plot reveals that the steepest $\Nu$ scaling range at low $\Ra$ \cite{Cheng2015,Cheng2020,Lu2021}, related to the range of the geostrophic regime where columns or cells are found \cite{Kunnen2021}, is pushed towards lower $\Ra$, out of reach for the current lower-$\Pr$ results but clearly observed in the data from \cite{Cheng2020}. Instead, a constant power law relation $\Nu=(0.035\pm0.008)\Ra^{(0.347\pm0.008)}$ can be used to describe the data well (weighted power-law fit, fit coefficients reported with $95\%$ confidence intervals), which is notably steeper than the non-rotating scaling $\Nu_{\red{0}}=0.11\Ra^{0.308}$ from \cite{Cheng2020}. For $\Ra\ge 2\times 10^{11}$, the two rotating datasets meet. Below that value, there is a significant difference in heat transfer between $\Pr=3.7$ and $\Pr=5.2$. This effect of rotation is remarkable; such a small change in Prandtl number certainly does not affect non-rotating convective heat transfer that much, by a few percent at most \cite{Ahlers2009}.

\red{Figure \ref{fig:Nu_vs_Ra}(b) shows the data normalized with the nonrotating $\Nu_0$ data from \cite{Cheng2020}, directly indicating the damping of convective heat transfer by rotation. Note that we use the same nonrotating $\Nu_0(\Ra)$ relation to compensate both rotating datasets (even though they are at different $\Pr$) as we do not have nonrotating reference data at $\Pr=3.7$. With only minute $\Pr$ effects expected for nonrotating convection (the Grossmann-Lohse model \cite{Ahlers2009} predicts less than $2\%$ difference) we do not introduce significant changes by this approach.} Additionally, \red{Fig. \ref{fig:Nu_vs_Ra}(b) shows that both datasets (at $\Pr=3.7$ and $5.2$)} do not \red{display} an `overshoot' of $\Nu$ above its non-rotating value $\red{\Nu_0}$, in line with the previous data at $\Pr=5.2$ \cite{Cheng2020}. This prominent effect for moderate $\Ra$ and $\Pr$ (e.g. Refs. \cite{Stevens2010,Yang2020,Anas2024}) indeed vanishes at large $\Ra$ and small $\Ek$ \cite{Cheng2015,Cheng2020}.

\subsection{Wall mode analysis\label{ch:wallmode}}
\begin{figure}
    \begin{tabular}[t]{llll}
    \raisebox{0.37\textwidth}{(a)} & \includegraphics[scale=0.57]{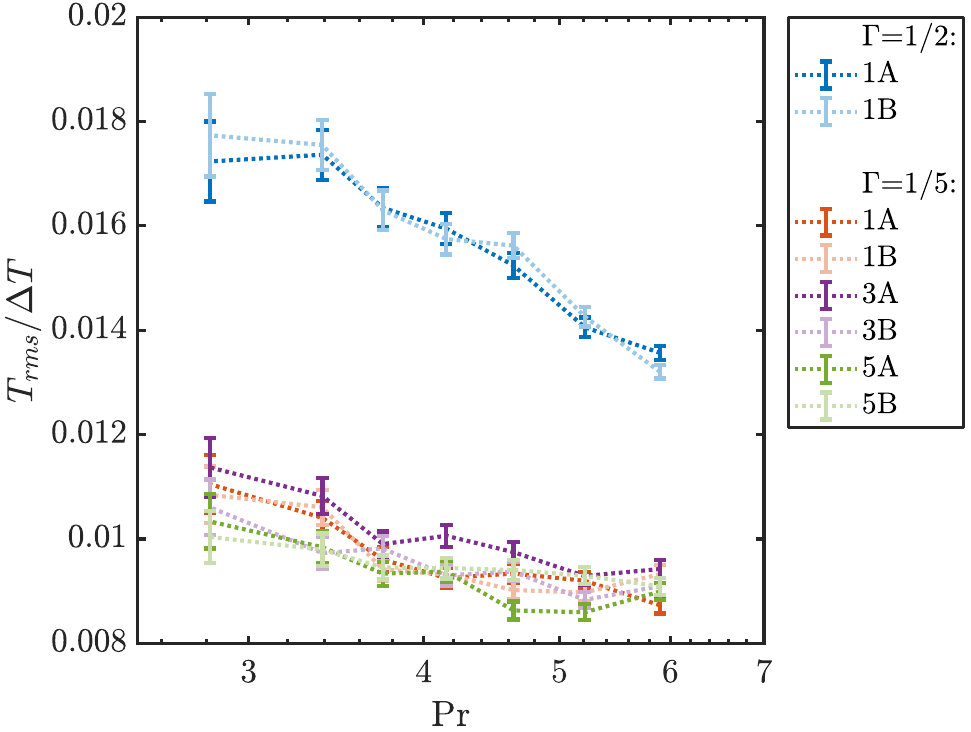} &
    \raisebox{0.37\textwidth}{(b)} & \includegraphics[scale=0.42]{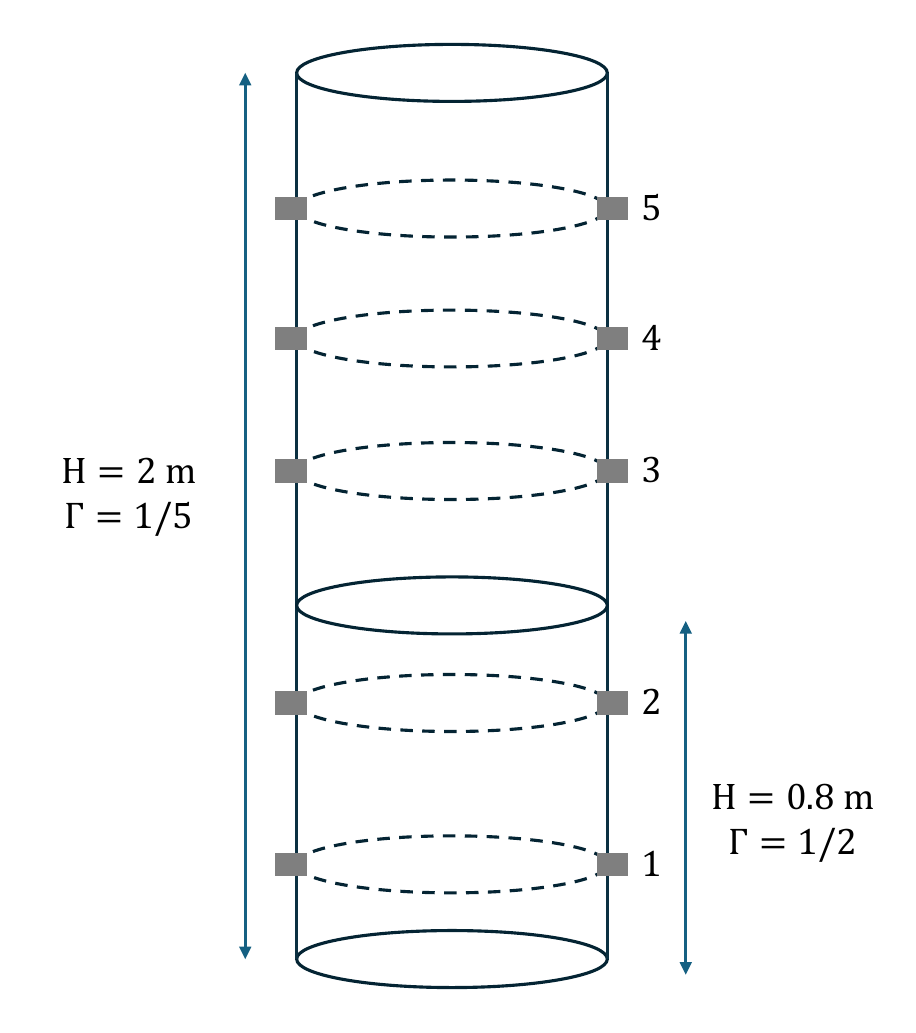}
    \end{tabular}
    \caption{\red{(a)} Normalized magnitude $T_{rms}/\Delta T$ of temperature fluctuations as measured by the sidewall temperature probes. Digits $1,3,5$ indicate increasing distance from the bottom plate while labels A, B are for laterally opposite probes at the same height. \red{(b) A schematic drawing of the setup with the sidewall sensors indicated in grey for both aspect ratios. The sensors were placed at midheight of each section of the cylinder.}}
    \label{fig:Trms_sidewall}
\end{figure}

In this section we use the sidewall temperature probes (see Sec. \ref{ch:meth} and also \cite{Cheng2020}) to analyze the fluctuations induced by the drifting wall mode: as the wall mode drifts azimuthally, a fixed probe will measure a quasi-periodic signal of alternating hot and cold excursions around a mean value. The amplitude of the oscillating signal can be considered as the `strength' of the wall mode. Here, we will consider the root-mean-square magnitude of fluctuations around the mean as a measure for the wall mode strength. The measured $T_{rms}$, normalized with $\Delta T$, are plotted as a function of $\Pr$ in figure \ref{fig:Trms_sidewall}. Several sidewall temperature probes are included, where the digit $1,3,5$ indicates position from the bottom (at $z=0.2,1.0,1.8$ m, respectively\red{, see also the sketch in figure \ref{fig:Trms_sidewall}(b)}). Letters A, B identify the probes on laterally opposite sides at the same height. Probe pairs 2A/2B and 4A/4B are left out of the plot. We found in post-processing that pair 2A/2B were not functioning well: they did not follow the fluctuations as all others did. We expect these were not in good thermal contact with the sidewall, that an air pocket acted as a filter and flattened the fluctuating temperature signal. Given that all other probes (including 4A/4B) gave basically the same result (see figure \ref{fig:Trms_sidewall}) we felt that including 4A/4B would only clutter the graph further. 

We can see that the relative magnitude of fluctuations is larger for the smaller tank ($\Gamma=1/2$, $H=0.8$ m) than for the larger tank ($\Gamma=1/5$, $H=2$ m). This is probably a Rayleigh number effect. While there is no direct reference data available for the temperature amplitude of the wall mode at these parameter values, Zhang et al. \cite{Zhang2024} found a normalized temperature amplitude that is gradually reducing as $\Ra$ increases within a range of $\Ra$ around the onset of bulk convection.

There is a gradual downward trend of $T_{rms}/\Delta T$ as a function of $\Pr$, or in other words, the wall mode is more prominent at lower $\Pr$. This trend is most obvious at $\Gamma=1/2$; for $\Gamma=1/5$ the trend is reduced in magnitude. To get to the lowest $\Pr=2.8$, the mean temperature is $T_m=61^\circ$C. The thermal expansion coefficient $\alpha$ becomes larger at higher temperature while kinematic viscosity goes down (see also Figure \ref{fig:waterproperties}). This means that a rather small temperature difference $\Delta T=1.9^\circ$C is enough to achieve the same constant $\Ra$. The intrinsic accuracy of the temperature control leads to a larger relative uncertainty for smaller $\Delta T$. That explains why the error bars at lower $\Pr$ are larger.

A final observation is the vertical consistency of $T_{rms}$: at all heights we observe the same temperature amplitude. This implies that, at least for what concerns temperature, the wall mode has a roughly constant amplitude independent of the vertical position.

Based on these observations we expect that the wall mode contributes more to the overall convective heat transfer (Nusselt number) at low $\Pr$ than at high $\Pr$. This is in line with the simulation results of Zhang et al. \cite{Zhang2021}, who reported a similar trend of decreasing relative wall mode contribution to $\Nu$ as $\Pr$ is increased. Nonetheless, comparing the dependence on $\Pr$ of $T_{rms}/\Delta$ and $\Nu$, we see that $T_{rms}/\Delta T$ reduces by a smaller factor than $\Nu$, so we expect that the changes in convective heat transfer are partially but not fully due to the wall mode; the bulk heat transfer should change, too.

\section{Conclusion\label{ch:concl}}
In this paper, we have measured the efficiency of convective heat transfer (Nusselt number $\Nu$) in rotating Rayleigh--B\'enard convection in the \red{transition range between rotation-affected and geostrophic convection}. We focus on the dependence on the Prandtl number $\Pr$, that we could vary between $\Pr=2.8$ and $6$ by using water at different mean temperatures. For two heights of the cylindrical convection cell, $H=0.8$ and $2$ m (aspect ratio $\Gamma=1/2$ and $1/5$), we kept $\Ra$ constant at $1\times 10^{11}$ and $1.1\times 10^{12}$, respectively. With increasing $\Pr$ in the specified range, keeping also the Ekman number $\Ek=3\times 10^{-7}$ constant, we observe at both heights a gradual but significant reduction of $\Nu$ by approximately $25\%$. We could achieve a decent collapse of the two $\Ra$ cases by plotting $\Nu/\Ra^{0.41}$. The observed Prandtl number effect is significantly larger than what is found in non-rotating convection, where, in this range of $\Pr$, $\Nu$ only changes by a few percent \cite{Ahlers2009}.

We expect that this significant dependence on $\Pr$ for geostrophic rotating convection is related to its rich flow phenomenology. There are subregions of parameter space for cells, Taylor columns, plumes and geostrophic turbulence \cite{Julien2012,Kunnen2021} each with their own heat transfer scaling. The $\Pr$ dependence of the boundaries between these subregions is basically unexplored; here we provide evidence that changing $\Pr$ appears to shift them considerably. This is further supported by a measurement series varying $\Ra$ at constant $\Pr=3.7$ which is lower than for our previous measurements at $\Pr=5.2$ \cite{Cheng2020}. Within the same $\Ra$ range, we see a clear change of $\Nu(\Ra)$ scaling for $\Pr=5.2$ (identified as the columns-to-plumes transition), while a single uniform scaling is found for $\Pr=3.7$. Hence, we expect that we could not reach down to the columns region at $\Pr=3.7$, a significant shift of the transition point.

The effect of the wall mode on convection is studied with the temperature probes in the sidewall. We interpret the normalized magnitude of temperature fluctuations $T_{rms}/\Delta T$ as a measure for the amplitude of the wall mode. The analysis reveals that the wall mode is more intense at lower $\Pr$, contributing a larger fraction of the total convective heat transfer than at higher $\Pr$, but not enough more to completely cover the $25\%$ reduction of $\Nu$. Thus the bulk convection must also change.

The dependence of rotating convection on the Prandtl number remains underexplored, particularly in the geostrophic regime. We have contributed the first experimental data for $2.8\le\Pr\le 6$. Ideally, this range should be extended, with most changes expected for lowering $\Pr$ further, towards values $\Pr\approx 0.7$ that are common for gases. Unfortunately, we were hitting some technical limitations of our current setup that prevented us from going to even higher mean working temperatures $T_m$. At the same time, even higher $T_m$ would lead to even smaller required temperature differences between top and bottom plate, which can lead to problems of measurement accuracy and temperature control. The use of different liquids is another way to extend the $\Pr$ range, although not the most convenient (nor the most economical) for a large vessel like this (its volume at $\Gamma=1/5$ is $239$ L). Another approach to tackle this problem is to use numerical simulation, where $\Pr$ is a free parameter. However, that comes with its own challenges: simulations in the geostrophic regime require quite extreme parameter values (large $\Ra\gtrsim 10^{10}$ and small $\Ek\lesssim 10^{-6}$) which\red{, while possible (e.g., Refs. \cite{Song2024scaling,Song2024}),} is computationally expensive.

The study of \red{rapidly rotating convection} remains interesting due to the richness of the flow phenomenology and profound dependence on the values of the governing parameters. Notwithstanding the operational challenges for both experiments and numerical simulations, progress is made step-by-step to elucidate the properties of this regime that is of great relevance for the understanding of geophysical and astrophysical flows.

\begin{acknowledgments}
This publication is part of the project `Universal critical transitions in constrained turbulent flows' with file number VI.C.232.026 of the research programme NWO-Vici which is financed by the Dutch Research Council (NWO).
\end{acknowledgments}

\appendix
\section{Experiment specifications}\label{app:experiments}
\begin{figure}[h]
    \includegraphics[scale=0.57]{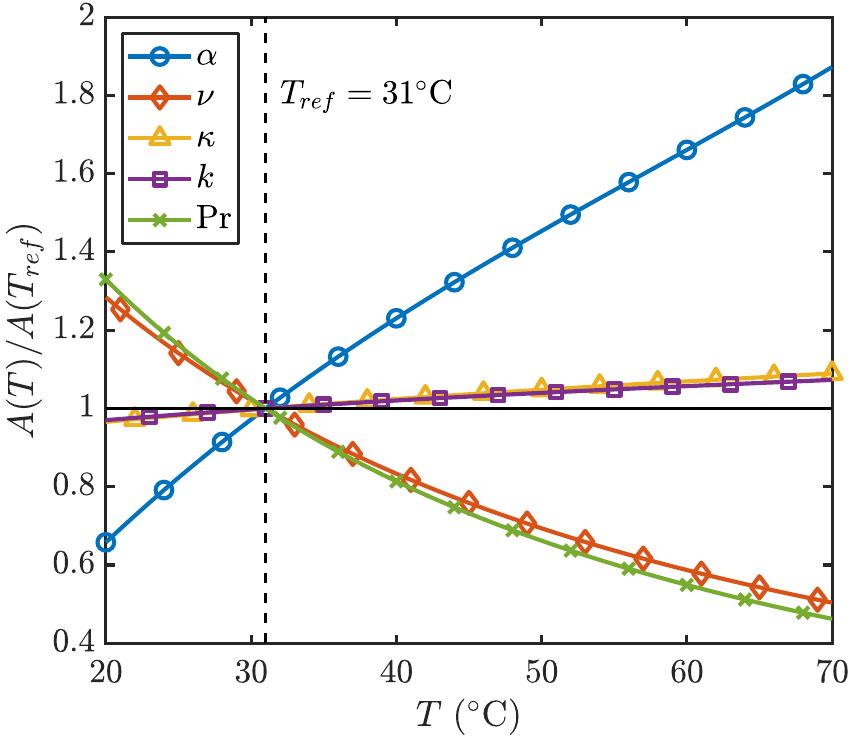}
    \caption{Plot of relevant water properties as a function of temperature: thermal expansion coefficient $\alpha$, kinematic viscosity $\nu$, thermal diffusivity $\kappa$, thermal conductivity $k$ and the Prandtl number $\Pr=\nu/\kappa$. They are plotted relative to the reference temperature $T_{ref}=31^\circ$C of our previous study \cite{Cheng2020}. Curves plotted according to polynomial relations given by Lide \cite{Lide2000}. At $T_{ref}=31^\circ$C, these give $\alpha=3.15\times 10^{-4}$ K$^{-1}$, $\nu=7.73\times 10^{-7}$ m$^2$/s, $\kappa=1.48\times 10^{-7}$ m$^2$/s, $k=0.616$ W/(m K) and $\Pr=5.22$.}
    \label{fig:waterproperties}
\end{figure}

This appendix lists all experimental conditions employed in this paper. Table \ref{tab:0.8prandtl} summarizes the measurements with variation of $\Pr$ at $\Gamma=1/2$. Table \ref{tab:2prandtl} does the same for $\Gamma=1/5$. Finally, Table \ref{tab:0.8rayleigh} contains the parameters for the $\Ra$ scan at constant $\Pr=3.7$. Additionally, we plot the temperature dependence of the relevant properties of water in Figure \ref{fig:waterproperties}.

\begin{table}[h]
\caption{Experiment specifications for the $\Gamma=1/2$ measurement series for varying $\Pr$.}
\label{tab:0.8prandtl}
\begin{tabular}{cccccccccc}
\hline
\hline
$\Gamma$ & $T_m$ ($^\circ$C)      & $\Delta T$ ($^\circ$C) & $\Pr$ & $\Ra$              & $\Ek$               & $\Ro$  & $\red{\widetilde{\Ra}}$ & $\Fr$ & $\Nu$      \\ 
\hline
0.494    & 26.03                  & 9.51                   & 5.90 & $9.85\times10^{10}$ & $3.10\times10^{-7}$ & 0.0401 & \red{206} & 0.099 & $214\pm 2$ \\
0.494    & 31.01                  & 7.68                   & 5.21 & $1.02\times10^{11}$ & $3.10\times10^{-7}$ & 0.0434 & \red{214} & 0.080 & $218\pm 3$ \\
0.494    & 36.02                  & 5.94                   & 4.64 & $9.80\times10^{10}$ & $3.10\times10^{-7}$ & 0.0451 & \red{206} & 0.065 & $219\pm 3$ \\
0.494    & 41.02                  & 4.93                   & 4.15 & $9.80\times10^{10}$ & $3.10\times10^{-7}$ & 0.0476 & \red{206} & 0.053 & $222\pm 4$ \\
0.494    & 46.00                  & 4.16                   & 3.74 & $9.78\times10^{10}$ & $3.10\times10^{-7}$ & 0.0501 & \red{205} & 0.044 & $231\pm 5$ \\
0.494    & 51.00                  & 3.56                   & 3.39 & $9.75\times10^{10}$ & $3.10\times10^{-7}$ & 0.0526 & \red{205} & 0.037 & $250\pm 7$ \\
0.494    & 60.99                  & 2.69                   & 2.81 & $9.70\times10^{10}$ & $3.11\times10^{-7}$ & 0.0557 & \red{204} & 0.026 & $275\pm 12$ \\
\hline                                
\hline
\end{tabular}
\end{table}

\begin{table}[h]
\caption{Experiment specifications for the $\Gamma=1/5$ measurement series for varying $\Pr$.}
\label{tab:2prandtl}
\begin{tabular}{cccccccccc}
\hline
\hline
$\Gamma$ & $T_m$ ($^\circ$C)      & $\Delta T$ ($^\circ$C) & $\Pr$ & $\Ra$              & $\Ek$               & $\Ro$ & $\red{\widetilde{\Ra}}$ & $\Fr$ & $\Nu$      \\
\hline
0.195    & 26.05                  & 6.71                   & 5.90 & $1.13\times10^{12}$ & $2.99\times10^{-7}$ & 0.131 & \red{$2.27\times 10^3$} & 0.003 & $568\pm10$ \\
0.195    & 31.02                  & 5.26                   & 5.21 & $1.14\times10^{12}$ & $2.99\times10^{-7}$ & 0.140 & \red{$2.29\times 10^3$} & 0.002 & $577\pm10$ \\
0.195    & 36.02                  & 4.23                   & 4.64 & $1.14\times10^{12}$ & $2.99\times10^{-7}$ & 0.148 & \red{$2.28\times 10^3$} & 0.002 & $589\pm12$ \\
0.195    & 41.01                  & 3.49                   & 4.15 & $1.13\times10^{12}$ & $3.00\times10^{-7}$ & 0.157 & \red{$2.27\times 10^3$} & 0.001 & $610\pm13$ \\
0.195    & 46.00                  & 2.94                   & 3.74 & $1.13\times10^{12}$ & $2.99\times10^{-7}$ & 0.164 & \red{$2.26\times 10^3$} & 0.001 & $634\pm15$ \\
0.195    & 50.99                  & 2.52                   & 3.39 & $1.12\times10^{12}$ & $2.99\times10^{-7}$ & 0.172 & \red{$2.24\times 10^3$} & 0.001 & $673\pm21$ \\
0.195    & 60.98                  & 1.90                   & 2.81 & $1.12\times10^{12}$ & $2.99\times10^{-7}$ & 0.189 & \red{$2.23\times 10^3$} & 0.001 & $759\pm38$ \\
\hline                       
\hline
\end{tabular}
\end{table}

\begin{table}[h]
\caption{Experiment specifications for the $\Gamma=1/2$ measurement series for varying $\Ra$.}
\label{tab:0.8rayleigh}
\begin{tabular}{cccccccccc}
\hline
\hline
$\Gamma$ & $T_m$ ($^\circ$C)      & $\Delta T$ ($^\circ$C) & $\Pr$ & $\Ra$              & $\Ek$               & $\Ro$  & $\red{\widetilde{\Ra}}$ & $\Fr$ & $\Nu$     \\
\hline
0.494    & 46.00                  & 1.96                   & 3.74 & $4.62\times10^{10}$ & $3.10\times10^{-7}$ & 0.0345 & \red{97.0} & 0.044 & $186\pm10$ \\
0.494    & 46.00                  & 4.16                   & 3.74 & $9.78\times10^{10}$ & $3.10\times10^{-7}$ & 0.0501 & \red{205} & 0.044 & $231\pm 5$ \\
0.494    & 46.02                  & 9.92                   & 3.74 & $2.33\times10^{11}$ & $3.10\times10^{-7}$ & 0.0774 & \red{490} & 0.044 & $309\pm 5$ \\
0.494    & 46.01                  & 14.93                  & 3.74 & $3.51\times10^{11}$ & $3.10\times10^{-7}$ & 0.0950 & \red{737} & 0.044 & $359\pm10$ \\
0.494    & 46.02                  & 19.93                  & 3.74 & $4.69\times10^{11}$ & $3.10\times10^{-7}$ & 0.110  & \red{984} & 0.044 & $399\pm 4$ \\
0.494    & 46.19                  & 29.59                  & 3.73 & $7.00\times10^{11}$ & $3.09\times10^{-7}$ & 0.134  & \red{1463} & 0.044 & $458\pm 5$ \\
\hline
\hline
\end{tabular}
\end{table}

\bibliographystyle{unsrt}
\bibliography{sources}

\end{document}